# Unidirectional coherent quasiparticles in the high-temperature rotational symmetry broken phase of $AV_3Sb_5$ kagome superconductors

Hong Li[1], He Zhao[1], Brenden R. Ortiz[2], Yuzki Oey[2], Ziqiang Wang[1], Stephen D. Wilson[2] and Ilija Zeljkovic[1,*]

**Affiliations:**

[1] Department of Physics, Boston College, Chestnut Hill, MA 02467, USA

[2] Materials Department, University of California Santa Barbara, Santa Barbara, California 93106, USA

* Corresponding author: ilija.zeljkovic@bc.edu

**Introductory paragraph:**

**Kagome metals $AV_3Sb_5$ [1] (where the $A$ can stand for K, Cs, or Rb) display a rich phase diagram of correlated electron states, including superconductivity [2–4] and density waves [5–7]. Within this landscape, recent experiments revealed signs of a transition below approximately 35 K attributed to an electronic nematic phase that spontaneously breaks rotational symmetry of the lattice [8]. Here, we show that rotational symmetry breaking initiates universally at a high temperature in these materials, toward the 2 x 2 charge density wave transition temperature. We do this via spectroscopic-imaging scanning tunneling microscopy and study atomic-scale signatures of electronic symmetry breaking across several materials in the $AV_3Sb_5$ family: $CsV_3Sb_5$, $KV_3Sb_5$ and Sn-doped $CsV_3Sb_5$. Below a significantly lower temperature of about 30 K, we measure quantum interference of quasiparticles, a key signature for the formation of a coherent electronic state. These quasiparticles display a pronounced unidirectional feature in reciprocal space that strengthens as the superconducting state is approached. Our experiments reveal that high-temperature rotation symmetry breaking and the charge ordering states are separated from the superconducting ground state by an intermediate-temperature regime with coherent unidirectional quasiparticles. This picture is phenomenologically different compared to that in high-temperature superconductors, shedding light on the complex nature of rotation symmetry breaking in $AV_3Sb_5$ kagome superconductors.**

**Main Text:**

Geometric frustration and non-trivial electronic band topology inherent to the Kagome lattice naturally favor the emergence of exotic quantum states. Vanadium-based kagome superconductors $AV_3Sb_5$ ($A$ = K, Cs, Rb) [1,2] have recently attracted enormous attention as the playground to experimentally [5–20] and theoretically [21–29] explore unconventional phenomena in kagome metals. This family of materials exhibits a rich electronic behavior, such as the emergence of various density waves [5–7,15], time-reversal symmetry breaking [11], potentially unconventional superconductivity [16,17,30], and the tendency towards unidirectionality [5–8,10,19]. Signatures of the unidirectional character include the formation of a $4a_0$ charge stripe order [6], two-fold transport [7,19,30], and anisotropic amplitudes of the 2 x 2 CDW order at low temperature [5,8,10]. Recent experiments using a combination of NMR, elasto-transport and scanning tunneling microscopy (STM) measurements, revealed signs of another electronic transition at $T \approx 35$ K in $CsV_3Sb_5$ thought to be associated with a purely electronic nematic phase [8] and a precursor to superconductivity. In this work, we use high-resolution spectroscopic-imaging scanning tunneling microscopy (SI-STM) to study the temperature-dependence of electronic symmetry breaking in a range of different materials in this family of kagome superconductors. We find that a substantial rotation symmetry breaking electronic signal in $AV_3Sb_5$ emerges universally at a much higher temperature than previously thought [8], suggesting that electronic transitions uncovered thus far occur within a system where rotation symmetry had already been reduced. Importantly, we detect the onset of additional rotational symmetry breaking electronic wave vectors below 30 K, related to a unidirectional electron scattering and interference arising from the V kagome bands. The surprising emergence of these coherent quasiparticles at the Fermi level below this intermediate temperature should in principle be detectable in charge transport measurements, which could account for the anomalous features in recent elasto-transport [8] and magneto-transport [31] measurements.

Single crystals of $AV_3Sb_5$ are characterized by a layered hexagonal crystal structure ($a=b$=5.4 Å, $c$=9 Å) composed of $V_3Sb_5$ slabs stacked between alkali layers [1,2] (Fig. 1a). The central plane contains a kagome network of V atoms spaced within a hexagonal lattice of Sb atoms. A common feature identified across all members of this kagome family is a CDW state whose in-plane component results in a $2a_0$ x $2a_0$ supercell ($T^* \approx$ 70-100 K [2,3,32–35]) that develops in the normal state, above the superconducting transition. We explore several different materials in the $AV_3Sb_5$ family: $CsV_3Sb_5$ ($T^*$= 94 K [2]), Sn-doped $CsV_3Sb_{5-x}Sn_x$ ($T^* \approx$ 70-90 K for different Sn content $x$ [34]) and $KV_3Sb_5$ ($T^*$=78 K [3]). Our SI-STM measurements are focused on the Sb-terminated surface, cleaved at low temperature in ultra-high vacuum and immediately placed into the STM head (Methods). The Sb-terminated surface is shown to be robust against reconstruction [5–7] and provides a direct window into the momentum-resolved structure of the V kagome bands [6], which is not possible on the alkali-terminated surface (Extended Data Figure 1). We start by characterizing the low-temperature state at 4.5 K, first measuring the most widely studied material in this family thus far, $CsV_3Sb_5$. STM topographs of $CsV_3Sb_5$ show a hexagonal atomic structure, the well-established $2a_0$ by $2a_0$ charge density wave (CDW) and a unidirectional electronic modulation related to the $4a_0$ charge stripe order (Fig. 1c,d). The two charge ordering states can be clearly discerned as well-defined wave vectors in the Fourier transform (FT) of the STM topograph (Fig. 1d). Consistent with previous measurements [5–7,15], we note that these wave vectors do not disperse as a function of energy (Fig. 1e), indicative of the charge ordering nature of the states.

While the $4a_0$ charge stripe order by itself already breaks the rotation symmetry, we note that the amplitudes of the $2a_0$ by $2a_0$ CDW peaks also exhibit a two-fold anisotropy at 4.5 K along the same direction (Fig. 1f). This is consistent with the energy-dependent analysis previously used to establish rotation symmetry breaking at 4.5 K in $KV_3Sb_5$ [10] and $CsV_3Sb_5$ [8]. We rule out tip artifacts in artificially creating this anisotropy by imaging a CDW domain boundary [10]. While previous experiments only characterized the energy-dependent anisotropy of the $2a_0$ by $2a_0$ CDW state at low-temperature [5,8,10], there is little information about how this anisotropy evolves with temperature. We proceed to investigate if the observed CDW anisotropy is tied to the existence of the $4a_0$ charge stripe order that onsets below about 50 K [6], or if it disappears at higher temperatures as we approach and cross the $4a_0$ charge order transition. As examining individual STM topographs does not provide a complete picture of the anisotropy of the CDW state [10], we perform the energy-resolved mapping of the $2a_0$ by $2a_0$ CDW amplitudes [10], which has not been achieved to-date at these high temperatures. We track an identical region of the sample from 4.5 K to 55 K. We first confirm that the $4a_0$ charge stripe order present in STM topographs at 4.5 K (Fig. 2a,c) is absent at 55 K (Fig. 2b,d). Our energy-dependent measurements reveal that the two-fold anisotropy in the peak amplitudes of the three $2a_0$ wave vectors remains qualitatively similar to that at low temperature – there exists one dominant direction that is noticeably different than the other two (Fig. 2g,h). The dominant wave vector is in the same direction in which the $4a_0$ order subsequently condenses at lower temperature. In real space, rotation symmetry breaking can be visualized by examining individual $dI/dV(\mathbf{r}, V)$ maps, in which noticeable unidirectionality can be clearly seen in the presence (Fig. 2e) or the absence of $4a_0$ charge order (Fig. 2f). Rotation symmetry breaking in this channel has been detected up to the highest accessible temperatures in this work of about 60 K. We note that while the anisotropy of CDW amplitudes between the three directions is reduced at a higher temperature, rotation symmetry breaking signal is still substantial (Fig. 2h); the peak amplitude at the dominant CDW wave vector only approximately a factor of 2 different between the two temperatures across a wide energy range (Fig. 2g,h).

We find that that the high-temperature $C_2$-symmetric CDW signal is pronounced and universal across all members of the $AV_3Sb_5$ family studied (Fig. 3). In both pristine $CsV_3Sb_5$ and a higher Sn concertation of $CsV_3Sb_{5-x}Sn_x$, two-fold symmetric CDW amplitudes are still apparent at 55 K, at a temperature higher than the formation temperature of the $4a_0$ charge order (Fig. 3a-d). The same is confirmed in $KV_3Sb_5$ where long-range $4a_0$ unidirectional charge order is notably absent in the entire temperature range (Fig. 3e,g). These experiments show a natural tendency of this family towards high-temperature rotation symmetry breaking associated with the CDW state, robust against the disorder introduced by Sn dopants.

Next, we turn to the most intriguing aspect of our temperature-dependent SI-STM measurements. Our previous work established that by imaging electron scattering and interference in STM $dI/dV(\mathbf{r}, V)$ maps of the Sb surface, we can gain direct insight into the electronic band structure of this system [6]. In particular, FTs of $dI/dV(\mathbf{r}, V)$ maps in $CsV_3Sb_5$ at 4.5 K show a remarkable two-fold scattering signature, tied to the renormalization of the V bands (Fig. 4a,b). These scattering wave vectors come from the quasi-1D sections of the V $d$ bands (inset in Fig. 4b, Extended Data Figure 2), and disperse with energy concomitant with the evolution of the band structure (Fig. 4d). By performing temperature-dependent SI-STM measurements, we find that this scattering signal gradually decreases in strength as the temperature is raised, and vanishes within our resolution approaching on approximately 30 K (Fig. 4c). Quantitatively, this can be seen by extracting the intensity of the scattering wave vector and plotting it as a function of temperature (Fig. 4e,f). To demonstrate that this is not accidental, we repeat the measurements on a Sn-doped $CsV_3Sb_5$ sample (Extended Data Figure 3). We find that the equivalent $C_2$-symmetric scattering wave vectors, which appear comparable to that in the pristine system, also disappear approximately at the same temperature (Fig. 4g). We repeated the equivalent measurement for $KV_3Sb_5$, and find that the $C_2$-symmetric electron scattering gets suppressed with increased temperature in a similar manner, with the expected complete suppression at a comparable temperature (Extended Data Figure 4). The small difference between the onsets of coherent quasiparticles in different members of $AV_3Sb_5$ (Fig. 4f,g, Extended Data Figure 4) should be further explored in future experiments in a wider doping range across both superconducting domes [34]. We stress that anisotropic electron scattering in Fig. 4 and the $4a_0$ charge stripe order in $CsV_3Sb_5$ and Sn-doped $CsV_3Sb_5$ do not appear to be intimately related. We can conclude this based on the difference in the onset temperatures of the $4a_0$ order and anisotropic QPI in pristine $CsV_3Sb_5$ and Sn-doped $CsV_3Sb_5$, as well as the observation of anisotropic scattering signature in $KV_3Sb_5$ where the long-range $4a_0$ order does not form down to at least 4.5 K (Extended Data Figure 5).

Our experiments reveal a complex landscape of temperature-driven phenomena in $A$V$_3$Sb$_5$. In contrast to recent experiments reporting that rotation symmetry breaking is nearly absent above 35 K based on STM topographic measurements on the Cs surface [8], our energy-resolved SI-STM measurements demonstrate that rotation symmetry breaking is substantial at a much higher temperature. We observed the same behavior on multiple CsV$_3$Sb$_5$ samples and different areas of the samples (Figure 3, Extended Data Figure 6a-d). We further show that it is distinct from the 4$a_0$ charge stripe order that subsequently forms at a lower temperature. We note that due to technical challenges of the measurements, we are currently unable to perform energy-resolved temperature-dependent measurements tracking the same region of the sample from 60 K up to the slightly higher $T^* \approx$ 70 K - 95 K in different members of the $A$V$_3$Sb$_5$ studied here. Given that experimental signatures of emergent phenomena appearing below $T^*$ in time-resolved spectroscopy [36,37], Raman scattering [38,39], X-ray diffraction [40], muon spin spectroscopy [41], elasto-transport [8], NMR measurements [8,42] and magneto-transport [31] are all detected at temperatures lower than 60-70 K, it is conceivable that the onset of rotation symmetry breaking clearly seen in our measurements up to at least 60 K persists up to $T^*$. In this scenario, both translation and rotation symmetry in $A$V$_3$Sb$_5$ are simultaneously broken at $T^*$. Consistent with this picture, a C$_2$-symmetric state that emerges at $T^*$ was recently also revealed by scanning birefringence microscopy [43]. The anisotropy in the 2 x 2 CDW state could be related to the 3D stacking order, where the anisotropic signal measured by STM is due to the coupling of neighboring kagome planes, where this interaction becomes increasingly stronger as the system is cooled down.

Importantly, we discover a distinct emergence of coherent quasiparticles substantially below $T^*$. This has been repeatedly observed on different CsV$_3$Sb$_5$ samples and other members of the $A$V$_3$Sb$_5$ family (Figure 4, Extended Data Figure 3, Extended Data Figure 4, Extended Data Figure 6e-h, Extended Data Figure 7, Extended Data Figure 8). Based on their reciprocal-space signature, these quasiparticles are related to the V electronic bands. As they are located at the Fermi level, charge transport measurements should naturally be sensitive to their existence and spatial symmetry. In fact, elasto-transport [8], magnetotransport [31] and thermoelectric [44] measurements recently revealed unusual features emerging concomitant with the onset of unidirectional quasiparticles observed here at about 30 K. Anomalous features below 30 K have also been reported in coherent phonon spectroscopy [37] and muon spin spectroscopy relaxation rates [41,45]. As such, our experiments may provide a microscopic insight into the origin of a peculiar intermediate transition in bulk measurements. Our work also demonstrates that superconductivity in $A$V$_3$Sb$_5$ develops from a normal state with coherent quasiparticles that break rotational symmetry, even in the absence of the 4$a_0$ unidirectional charge order observed on the Sb terminated surface in CsV$_3$Sb$_5$, and persist below superconducting $T_c$. This in turn raises a possibility that $A$V$_3$Sb$_5$ may have a nematic superconducting ground state. Lastly, unidirectional electron scattering and interference in high-temperature superconductors has been viewed as a smoking gun signature of electronic nematicity [46,47] and it is generally believed that such unidirectional scattering marks the onset of rotational symmetry breaking of the electronic structure. In contrast to this picture of rotational symmetry breaking in high-temperature superconductors, we show that unidirectional quasiparticles in $A$V$_3$Sb$_5$ emerge far below the temperature where rotational symmetry of the electronic structure is first reduced. This highlights the complexity of phenomena associated with rotational symmetry breaking in $A$V$_3$Sb$_5$ distinct from that in high-temperature superconductors and sheds fresh light on the emergence of electronic nematicity in this family of kagome superconductors.

**Acknowledgements**

We are thankful for insightful conversations with Leon Balents, Rafael Fernandes, Liang Wu and Riccardo Comin. I.Z. gratefully acknowledges the support from NSF-DMR 2216080. S.D.W. and B.R.O. acknowledge support via the UC Santa Barbara NSF Quantum Foundry funded via the Q-AMASE-i program under award DMR-1906325. Z.W. acknowledges the support of U.S. Department of Energy, Basic Energy Sciences Grant No. DE-FG02-99ER45747 and the Cottrell SEED Award No. 27856 from Research Corporation for Science Advancement.

**Author Contributions**

STM experiments and data analysis were performed by H.L. and H.Z under the supervision of I.Z.. B.R.O. and Y.O. synthesized and characterized the samples under the supervision of S.D.W. Z.W. provided theoretical input. H.L., H.Z., S.D.W., Z.W. and I.Z. wrote the paper, with the input from all authors. I.Z. supervised the project.

**Competing Interests**

The Authors declare no Competing Financial or Non-Financial Interests.

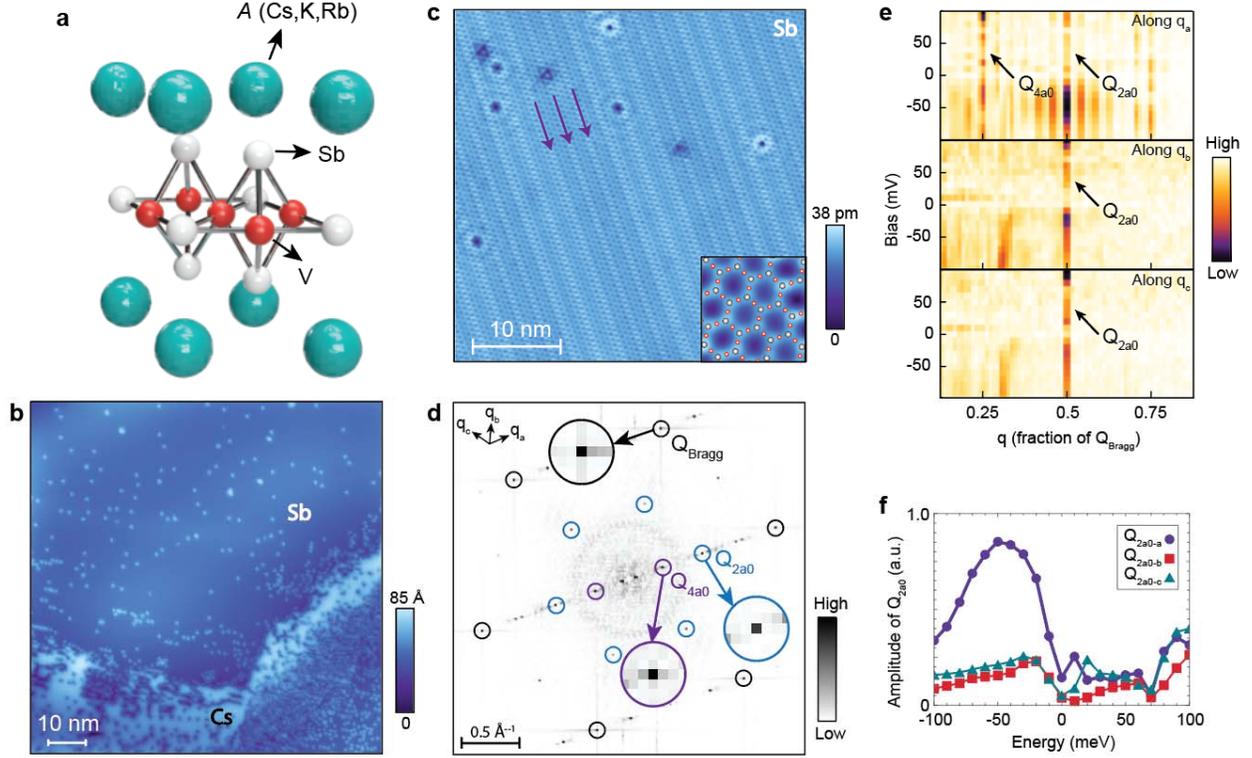

**Figure 1. Low-temperature electronic state of CsV$_3$Sb$_5$. (a)** A ball model of the unit cell of $A$V$_3$Sb$_5$ ($A$=Cs, K or Rb). Green, white and red spheres denote $A$, Sb and V atoms, respectively. **(b)** Large-scale STM topograph taken across a Cs-Sb step. The higher step is Cs termination and the lower one is the complete Sb surface. **(c)** STM topograph taken on the Sb termination of CsV$_3$Sb$_5$ showing the 4$a_0$ charge stripe modulation. The bottom right corner inset is a zoomed-in atomically-resolved topograph with the Sb-V ball model superimposed. **(d)** Fourier transform (FT) of the topograph in (c). Atomic Bragg peaks are enclosed in black circles, while the **Q**$_{2a0}$ and **Q**$_{4a0}$ charge ordering peaks are circled in blue and purple. The three insets demonstrate the sharpness of the FT peaks. **(e)** Waterfall plot of an FT linecut along **q**$_a$, **q**$_b$ and **q**$_c$ lattice directions as a function of energy, showing energy-independent reciprocal space position of **Q**$_{2a0}$ and **Q**$_{4a0}$ peaks. **(f)** FT intensities of the 2 x 2 CDW order peaks as a function of energy, similarly to the analysis in Ref. [14], showing one direction that is distinctly different compared to the other two. Purple circles, red squares and green triangles represent FT amplitudes of the **Q**$_{2a0-a}$, **Q**$_{2a0-b}$ and **Q**$_{2a0-c}$, respectively. STM setup conditions: (b) V$_{sample}$ = 200 mV, I$_{set}$ = 10 pA; (c) V$_{sample}$ = 100 mV, I$_{set}$ = 600 pA; (e,f) V$_{sample}$ = 100 mV, V$_{exc}$ = 4 mV, I$_{set}$ = 600 pA.

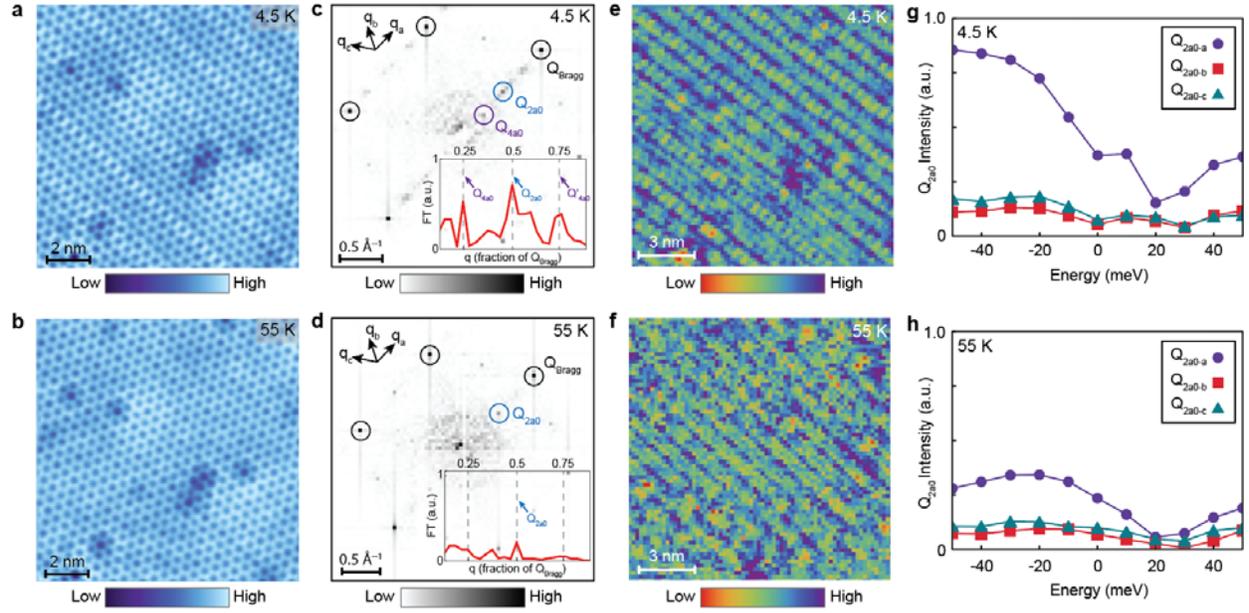

**Figure 2. High-temperature rotation symmetry breaking before the condensation of the 4$a_0$ charge stripe order. (a,b)** STM topograph of an identical region of the Sb surface of CsV$_3$Sb$_{4.98}$Sn$_{0.02}$ taken at (a) 4.5 K and (b) 55 K. **(c,d)** Fourier transforms (FTs) of (a,b). The bottom right inset in (c,d) shows an FT linecut from the center of the FT to the atomic Bragg peak **Q**$_{Bragg-a}$ along the **Q**$_{4a0}$ wave vector. **Q**$_{4a0}$ modulation is notably absent in (b,d). **(e,f)** d$I$/d$V$(**r**, $V$=-20 mV) taken over the region in (a,b) at (e) 4.5 K and (f) 55 K. **(g,h)** FT amplitudes of the three Q$_{2a0}$ peaks at (g) 4.5 K and (h) 55 K as a function of energy, showing that the two-fold anisotropy persists even after the 4$a_0$ charge order disappears at high temperature. STM setup conditions: (a,b) V$_{sample}$ = 50 mV, I$_{set}$ = 100 pA; (e,f) V$_{sample}$ = 50 mV, V$_{exc}$ = 10 mV, I$_{set}$ = 100 pA.

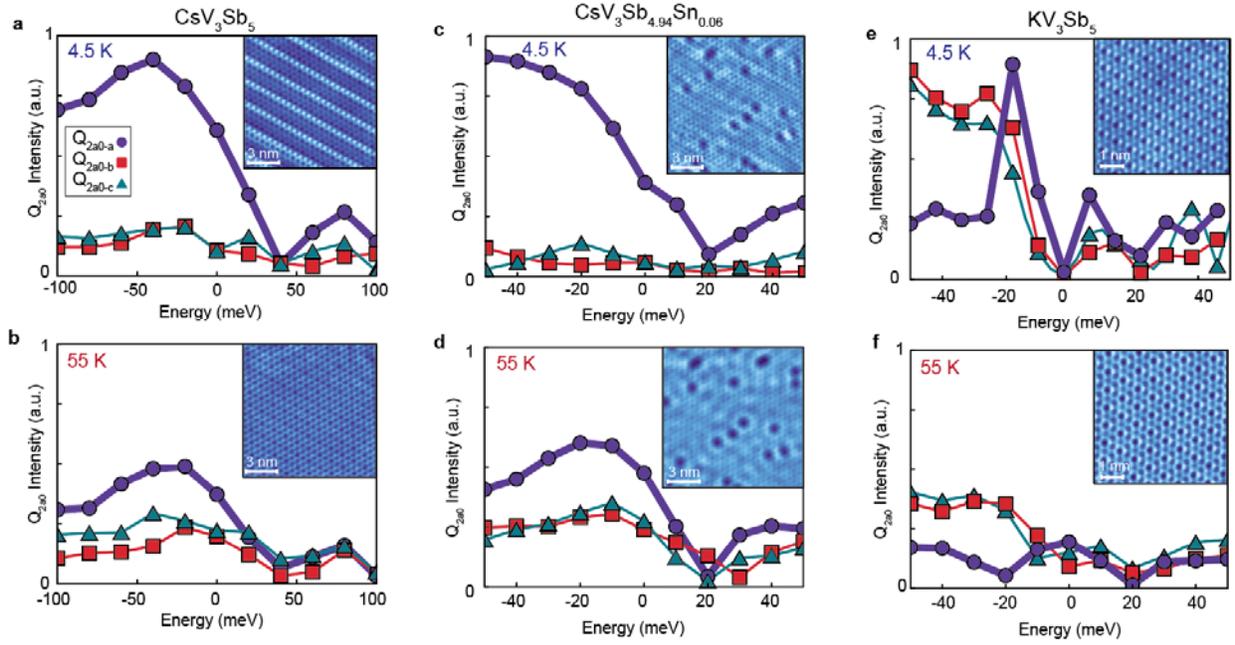

**Figure 3. Universality of the high-temperature C₂-symmetric signature across different members of the $AV_3Sb_5$ family. (a,b)** Fourier transform (FT) amplitudes of the 2 x 2 CDW order peaks as a function of energy from d$I$/d$V$ maps acquired on the Sb surface of $CsV_3Sb_5$ at (a) 4.5 K and (b) 55 K. Insets in (a,b) show STM topographs taken over the same region of the Sb-terminated surface at the corresponding temperatures. Equivalent panels to those in (a-b) for: **(c,d)** $CsV_3Sb_{4.94}Sn_{0.06}$ and **(e,f)** $KV_3Sb_5$. STM setup conditions: (a,b) Inset: $V_{sample}$ = 50 mV, $I_{set}$ = 200 pA; dI/dV maps: $V_{sample}$ = 100 mV, $V_{exc}$ = 20 mV, $I_{set}$ = 100 pA. (c,d) Inset: $V_{sample}$ = 50 mV, $I_{set}$ = 200 pA; dI/dV maps: $V_{sample}$ = 50 mV, $V_{exc}$ = 10 mV, $I_{set}$ = 200 pA. (e,f) Inset: $V_{sample}$ = 50 mV, $I_{set}$ = 400 pA; dI/dV maps: $V_{sample}$ = 50 mV, $V_{exc}$ = 4 mV, $I_{set}$ = 400 pA.

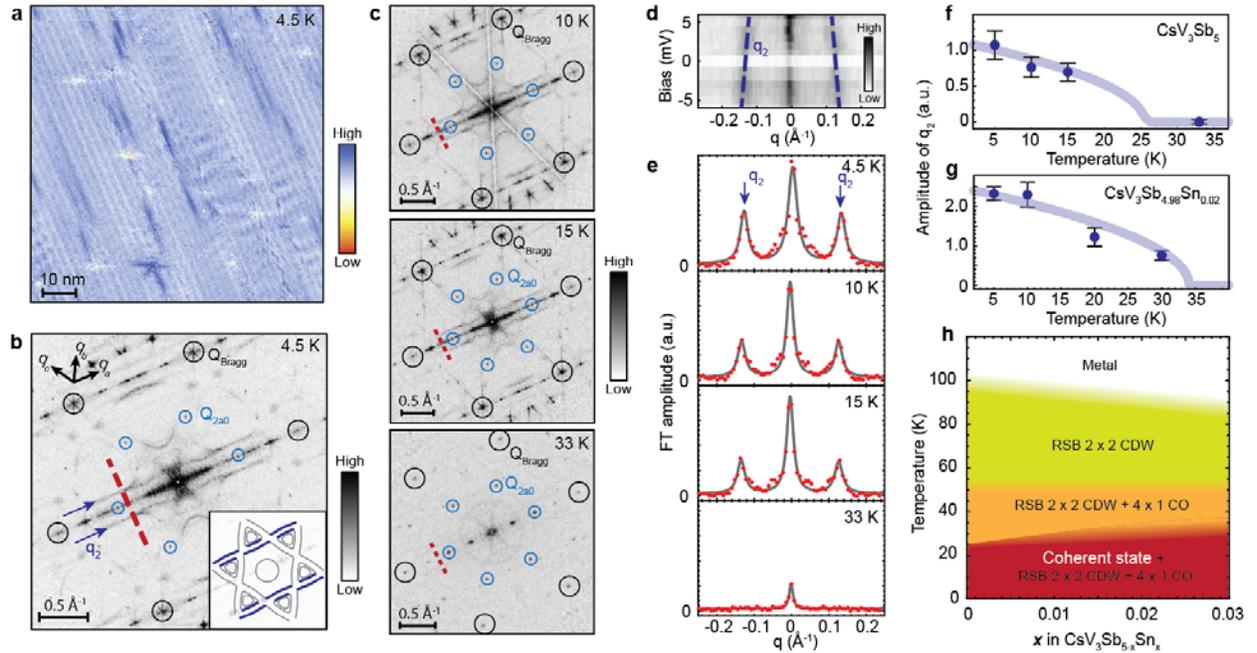

**Figure 4. Emergence of quasi-1D electron scattering below intermediate temperature of about 30 K.**
(**a**) A d$I$/d$V$(**r**, $V$=-4 mV) map taken over an 85 nm square region of the Sb surface of CsV$_3$Sb$_5$ at 4.5 K, and (**b**) its associated two-fold symmetrized Fourier transform (FT). Atomic Bragg peaks are enclosed in black circles, while blue arrows point along quasi-1D scattering wave vector **q**$_2$ originating from scattering from the parts of V bands. Inset in (b) shows a simplified schematic of the constant energy contour, with the relevant bands giving rise to the unidirectional scattering in STM FTs highlighted in thicker blue color. (**c**) Temperature-dependent two-fold symmetrized FTs showing the gradual disappearance of scattering features. For visual purposes, the center vertical line in the FT at 10 K in (c) was suppressed by subtracting the 2$^{nd}$ polynomial fit from each row in the raw dI/dV map before the data was two-fold symmetrized. (**d**) Energy-dependent FT linecut shown as a waterfall plot acquired along the red dashed line in (b) at 4.5 K, demonstrating the dispersive nature of **q**$_2$. Wave vector **q**$_2$ is detectable within about +/- 15 meV around the Fermi level. (**e**) Temperature-dependent FT linecut along the red dashed lines in (b,c), normalized by subtracting the FT background amplitude (see also Supplementary Figure 8). We use 0.36 Å$^{-1}$ average transverse to and 0.02 Å$^{-1}$ average along the linecut direction to create curves shown in (e). Gray lines in (e) represent Lorentzian fits to experimental data (red circles). Amplitude of **q**$_2$ as a function of temperature for (**f**) CsV$_3$Sb$_5$ and (**g**) CsV$_3$Sb$_{4.98}$Sn$_{0.02}$. Error bars in (f,g) represent standard errors obtained by fits to data shown in (e) and Supplementary Figure 3e. (**h**) A schematic phase diagram denoting the onset of different states as a function of doping in CsV$_3$Sb$_5$: coherent unidirectional state (measured here), 4a$_0$ charge order (CO) (ref. [4] and Supplementary Figure 11) and rotation symmetry broken (RSB) 2 x 2 CDW (Ref. [33]). STM setup conditions: (a-d) V$_{sample}$ = -6 mV, V$_{exc}$ = 1 mV, I$_{set}$ = 80 pA.

**Methods**

Bulk single crystals of $KV_3Sb_5$, $CsV_3Sb_5$ and $CsV_3Sb_{5-x}Sn_x$ were grown as described in Refs. [3,34]. To estimate the Sn content $x$ of $CsV_3Sb_{5-x}Sn_x$ samples, we acquire large-scale topographs of the Sb termination (around 50 nm or larger) and count the density of Sn dopants visible as dark hexagonal features in the STM topograph. $T^*$ for each $CsV_3Sb_{5-x}Sn_x$ sample is measured and reported in Ref. [34]. We cleave each sample at a cryogenic temperature as described in detail in Ref. [6] and immediately insert it into the STM head held at 4.5 K. All STM data was acquired using a customized Unisoku USM1300 microscope. Spectroscopic measurements were made using a standard lock-in technique with 915 Hz frequency and bias excitation as detailed in figure captions. STM tips used were home-made chemically-etched tungsten tips, annealed in a UHV chamber to a bright orange color before STM experiments. We apply the Lawler-Fujita drift-correction algorithm to all our data to align the atomic Bragg peaks onto single pixels with coordinates that are even integers. Representative unsymmetrized Fourier transforms are shown in Extended Data Figure 10.

**Data Availability**

The data used for analysis can be found on https://doi.org/10.5281/zenodo.7388403. All other data supporting the findings of this study are available upon request from the corresponding author.

**Code availability**

The computer code used for data analysis is available upon request from the corresponding author.